\documentclass[aps,pra,a4paper]{revtex4-1}
\usepackage{newtxtext,newtxmath,braket}
\usepackage[pdftex]{graphicx}
\begin{document}
\title{Quantum annealing with capacitive-shunted flux qubits}
\author{Yuichiro Matsuzaki$^{1}$}\email{matsuzaki.yuichiro@aist.go.jp}\author{Hideaki Hakoshima$^{1}$}\email{hakoshima-hideaki@aist.go.jp}\author{Yuya Seki$^{1}$}\author{Shiro Kawabata$^{1}$}
\affiliation{$^{1}$ Nanoelectronics Research Institute (NeRI), National Institute of Advanced Industrial Science and Technology (AIST)
Umezono 1-1-1, Tsukuba, Ibaraki 305-8568 Japan \\
}

\begin{abstract}
Quantum annealing (QA) provides us with a way to solve combinatorial optimization problems.
In the previous demonstration of the QA, a superconducting flux qubit (FQ) was used.  However, the flux qubits in these demonstrations have a short coherence time
such as tens of nano seconds. For the purpose to utilize quantum properties, it is necessary to use another qubit with a better coherence time. Here, we propose to use a capacitive-shunted flux qubit (CSFQ)
for the implementation of the QA. The CSFQ has a few order of magnitude better coherence time than the FQ used in the QA. We theoretically show that, although it is difficult to perform the conventional QA with the CSFQ due to the form and 
strength of the
interaction between the CSFQs, a spin-lock based QA can be implemented with the CSFQ even with the current technology. Our results pave the way for the realization of the practical QA that exploits quantum advantage with long-lived qubits.
\end{abstract}
\maketitle

\section{Introduction}
Quantum annealing (QA) and adiabatic quantum computation (AQC) are attractive ways to solve combinatorial optimization problems where a cost function is minimized 
with a suitable parameter set \cite{kadowaki1998quantum,farhi2000quantum,farhi2001quantum,albash2018adiabatic}.
It is known that we can map some optimization problems into a task of a ground-state search of the Ising Hamiltonian 
\cite{lucas2014ising}. 
Quantum annealing and AQC can find the ground state of the Ising Hamiltonian
by taking advantage of quantum fluctuations \cite{morita2008mathematical}.
In particular, AQC is guaranteed to converge to the ground state if the quantum adiabatic condition is satisfied \cite{albash2018adiabatic}.
There are many experimental and theoretical studies along this direction \cite{santoro2002theory,santoro2006optimization,lanting2014entanglement,boixo2014evidence,goto2016bifurcation}.

A superconducting flux qubit (FQ) \cite{mooij1999josephson,orlando1999superconducting} 
has been used to demonstrate the QA in the previous demonstration \cite{johnson2011quantum}.
The superconducting qubit is an artificial atom, and so we have many degrees of freedom to control the parameters \cite{makhlin2001quantum,clarke2008superconducting}.
The changes in the design of the superconducting circuit allow us to control the properties of the superconducting qubit in the fabrication.
After the fabrication, control lines coupled with the superconducting qubits can tune the parameters such as 
the qubit resonant frequency \cite{paauw2009tuning,zhu2010coherent}
and the coupling strength between qubits \cite{niskanen2007quantum}.
Such a controllability is essential for the implementation of the QA.
A potential problem is that  the FQs used in the QA have a short coherence time such as tens of nano seconds \cite{ozfidan2019demonstration}.
Such a short coherence time could make it difficult to exploit the quantum advantage for the QA.
There are many ways to improve the coherence time of the FQ.
Symmetric designs of the device is known to be one of the important ingredients for the long coherence time 
\cite{burkard2005asymmetry,bertet2005dephasing,yoshihara2006decoherence,bylander2011noise}.
The use of the 3D cavity containing the FQ can improve the coherence time due to the change in the environment of the FQ 
\cite{stern2014flux}. However, from the view point of the scalability and reproducability, 
further improvement is required to use the FQ for the QA to solve the practical problems with high coherence.

Recently, a capacitive shunt flux qubit (CSFQ) was fabricated 
\cite{you2007low,yan2016flux,corcoles2011protecting,steffen2010high} 
where
an additional capacitor shunted in parallel to the Josephson junction in the loop is introduced. This structure is known to suppress noise.
At the optimal point, the CSFQ has a coherence time of tens of micro seconds \cite{you2007low,yan2016flux}, which is a few order of magnitude 
longer than the FQs used in the QA.
Due to this long coherence time, the CSFQ is considered as a promising device to realize quantum information processing.

However, there are two main difficulties to use the CSFQ for the QA.
Firstly, the CSFQ has a much smaller coupling strength than the FQ used in the QA. The CSFQ has a persistent current of tens of nano 
ampere \cite{yan2016flux} while the FQs used in the QA has a few micro ampere of the persistent current \cite{johnson2011quantum}.
This results in a few orders of magnitude smaller coupling strength of the CSFQs than that of the FQs.
In the QA, it is necessary for the qubits to have a coupling strength that is comparable with the qubit resonant frequency, and the CSFQ may not satisfy such a condition due to the weak coupling strength.
Secondly, near the optimal operating point where the CSFQ has a longer coherence time, there are not only Ising type interaction 
but also a flip-flop type interaction with the CSFQ \cite{yan2016flux}.
In the QA, only Ising type interaction is required at the end of the calculation, and the residual flip-flop interaction could reduce the success probability to obtain an accurate solution. 

Here, we propose to implement the QA with the CSFQs by using a spin-lock technique.
The spin lock technique is a way to keep the quantum state in $\ket{+}$ (an eigenstate of $\sigma _x$) in a rotating frame.
Firstly, we prepare a state of $\ket{0}$ (a ground state of $\sigma _z$). Secondly, we perform a $\pi /2$ pulse of the microwave along the $y$ axis to prepare a state of $\ket{+}$.
Thirdly, we continuously perform the microwave driving along $x$ axis. Since $\ket{+}$ is an eigenstate of the Hamiltonian in the rotating frame, we can keep the state in $\ket{+}$ due to the driving, which is called the spin lock. 
This is a common technique in the field of the magnetic resonance \cite{loretz2013radio}.
Importantly, during the spin lock, the effective frequency of the qubit is the detuning between the qubit bare frequency and microwave driving frequency.
Also, if there is a large detuning between the qubit bare frequencies, 
the effect of the flip-flop type interaction between the qubits can be suppressed.
Therefore, the spin lock technique can overcome the limitation of the CSFQ to be used for the QA.
In the NMR, the spin-lock based QA has been discussed and demonstrated \cite{chen2011experimental,nakahara2013lectures}. While DC magnetic fields are applied along 
$x$ axis in the conventional QA, the AC driving fields along $x$ axis are applied in the spin lock-based QA.
However, since it is difficult to scale up in the NMR, a practical advantage is unclear to use the NMR for the QA. On the other hand, we propose to use the CSFQs for the spin-lock 
based QA, which is considered as a scalable device in quantum 
information processing.

We theoretically study the performance of the spin-lock based QA with the CSFQs.
The spin-lock based QA becomes equivalent to the conventional QA
under the assumption that the rotating wave approximation (RWA) is valid.
This means that, the performance of the spin-lock based QA can be degraded if the Rabi frequency of the driving is comparable with the qubit frequency.
Since the qubit frequency (an order of GHz) is just a few orders of magnitude larger than the other typical frequencies (an order of MHz) \cite{chen2011experimental}, a careful assessment of the error accumulation due to the violation of the RWA is necessary to evaluate 
the practicality of the spin-lock based QA with the CSFQs.
It is worth mentioning that, in our previous study at the SSDM \cite{matsuzakissdm}, we implement a numerical simulation along this direction with a fixed number of the qubits. 
On the other hand, we investigate how the performance of the QA with spin lock changes when we increase the number of the qubits, and discuss how our scheme scales in this paper. 
Moreover, we analyze the expression of the violation of the RWA.


\section{The conventional quantum annealing with DC transverse magnetic fields}
Let us review the conventional quantum annealing where DC transverse magnetic fields are applied \cite{kadowaki1998quantum}.
The Hamiltonian in the QA is described as follows.
\begin{align}
 H_{\rm{QA}}&=e^{-\gamma ^2t^2}H_{\rm{TR}}+(1-e^{-\gamma ^2t^2})H_{\rm{Ising}}, \\
H_{\rm{Ising}}&=-\sum_{i=1}^{L}\frac{h_i}{2} \sigma _z ^{(i)}-\sum_{i,i'=1}^{L}\frac{J_{i,i'}}{2} \sigma _z ^{(i)}\sigma _z ^{(i')}, \\
H_{\rm{TR}}&=-\sum_{i=1}^{L}\frac{\Lambda }{2} \sigma _x ^{(i)},
\end{align}
where $\gamma $ is the sweeping rate of the Hamiltonian, $h_i$ is the resonant frequency of the $i$-th qubits, $J_{i,i'}$ is the interaction strength between the $i$-th qubit and $i'$-th qubit, and 
$\Lambda $
denotes the amplitude of the transverse magnetic fields.
Here, we consider that the ground state of $H_{\rm{Ising}}$ is the solution of optimization problems which we want to find.

We first prepare a state of $\ket{++\cdots +}$, which is the ground state of $H_{\rm{QA}}$ at $t=0$, and we gradually turn off the transverse magnetic fields with a time scale determined by $\gamma$, while we adiabatically increase the term $H_{\rm{Ising}}$ with the same time scale. 
If $\gamma$ is much smaller than the energy gap between the ground state and first excited state of $H_{\rm{QA}}$ for all $t$, the quantum state continues the ground state of $H_{\rm{QA}}$ for all $t$ because of the quantum adiabatic theorem, and after the long time evolution ($t\gg \gamma^{-1}$), we can obtain a ground state of $H_{\rm{Ising}}$.

\section{QA with spin lock technique}
We explain the details of the QA with the spin-lock technique. 
Firstly, we prepare a state of $\ket{00\cdots 0}$. 
Secondly, we apply a global $\pi /2$ pulse so as to prepare a state of $\ket{++\cdots+}$. 
Thirdly, we continue driving all these qubits with the AC magnetic fields along $x$ direction, and gradually turn off the transverse driving fields while we gradually turn on $H_{\rm{Ising}}$. 
Finally, we measure the qubits. 
In the third step of this protocol, the state is governed by the unitary time evolution determined by the Hamiltonian $H$
\begin{align}
H&=H_0+e^{-\gamma ^2t^2}H_{\rm{D}}+(1-e^{-\gamma ^2t^2})(H_{\rm{Ising}}+H_{\rm{xy}}),\label{achamiltonian}\\
H_{0}&=\sum_{i=1}^{L}\frac{\omega +\delta\omega_i}{2} \sigma _z ^{(i)},\\
H_{\rm{D}}&= -\sum_{i=1}^{L}\lambda \cos{[(\omega+\delta\omega_i)t]} \sigma _x ^{(i)},\\
H_{\rm{xy}}&=-\sum_{i,i'=1}^{L}\frac{J_{i,i'}}{2} \sigma _x ^{(i)}\sigma _x ^{(i')}-\sum_{i,i'=1}^{L}\frac{J_{i,i'}}{2} \sigma _y ^{(i)}\sigma _y ^{(i')},
\end{align}
where $(\omega+\delta\omega_i)$ denotes a bare frequency of the $i$-th qubit and $\lambda$ denotes the Rabi frequency of the driving fields. 

In a rotating frame by the unitary transformation $\exp{[iH_0t]}$, we neglect the fast oscillating terms with the frequencies $(\omega+\delta\omega_i)$ in the Hamiltonian, which is called a rotating wave approximation (RWA).
Under this approximation, the Hamiltonian $H$ becomes equivalent to $H_{\rm{QA}}$ in the Eq.\ (1) by setting $\lambda=\Lambda$
This approximation becomes valid when $(\omega+\delta\omega_i)$ is sufficiently large.

\section{Evaluation of the performance of the  QA with the spin-lock technique}\label{numerical}
We investigate the potential errors during the spin-lock based QA derived from the violation of the RWA using numerical simulations. 
We solve a time-dependent Schr\"{o}dinger equation with Hamiltonian (4) from $t=0$ (the initial state is $\ket{++\cdots +}$) to $t = T_{{\rm end}}$. 
We demonstrate a ferromagnetic one-dimensional Ising model with free boundary condition. 
In particular, we consider the homogeneous case $h_j=h>0$ and $J_{i,i+1}=J>0$ $(i=1,2, \cdots , L-1)$ with the nearest neighbor interaction. 
In this case, the ground state of $H_{\rm{Ising}}$ is an all up state of $\ket{11\cdots 1}$. 
To quantify the accuracy to find the ground state of $H_{\rm{Ising}}$, we investigate a fidelity $F(t)=|\braket{11\cdots1|\Phi(t)}|^2$, where $\ket{\Phi(t)}$ denotes a solution of the Schr\"{o}dinger equation at a time $t$. 
$F(t)$ means the overlap between the true ground state of $H_{\rm{Ising}}$ and $\ket{\Phi(t)}$, which is the state created by the spin-lock based QA. 
If $F(t)$ is close to 1, $\ket{\Phi(t)}$ is close to the true ground state, and therefore our method works very well.

\begin{figure}
\includegraphics[width=15.0cm, clip]{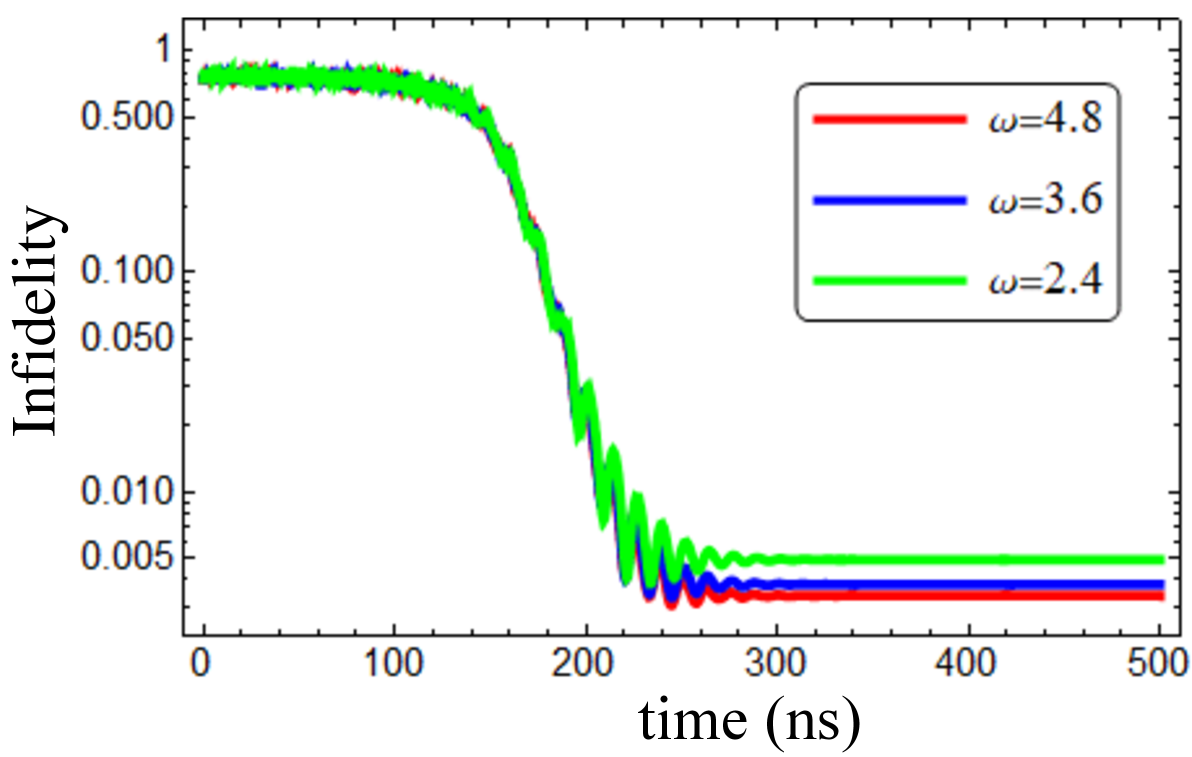}
\caption{The infidelity of the state plotted against time. 
Here, we set the parameters as $L=4$, $\lambda /2\pi =1$ GHz, $h/2\pi =0.03$ GHz, $\gamma =0.01$ GHz, $J/2\pi =0.05$ GHz, $T_{{\rm end}}=500$ ns, 
$\omega /2\pi =2.4, 3.6, 4.8$ GHz, and $\delta \omega _j /2\pi =1.9(j-1)$ GHz.
These parameters are typical in the CSFQ \cite{weber2017coherent}.
}
\label{f1}
\end{figure}

We plot an infidelity ($1-F(t)$) against time with feasible parameters of the CSFQs in the Fig. \ref{f1}. 
From this figure, we find that in all three cases, the infidelities are sufficiently close to 0 (less than 0.01), and we can achieve the high fidelity even by realistic parameters.
Moreover, the infidelity decreases as the qubit frequency increases. 
This reveals that the larger frequency of the qubit becomes the high accuracy of the RWA, which reduces the error in the spin-lock based QA.

\begin{figure}
\includegraphics[width=15.0cm, clip]{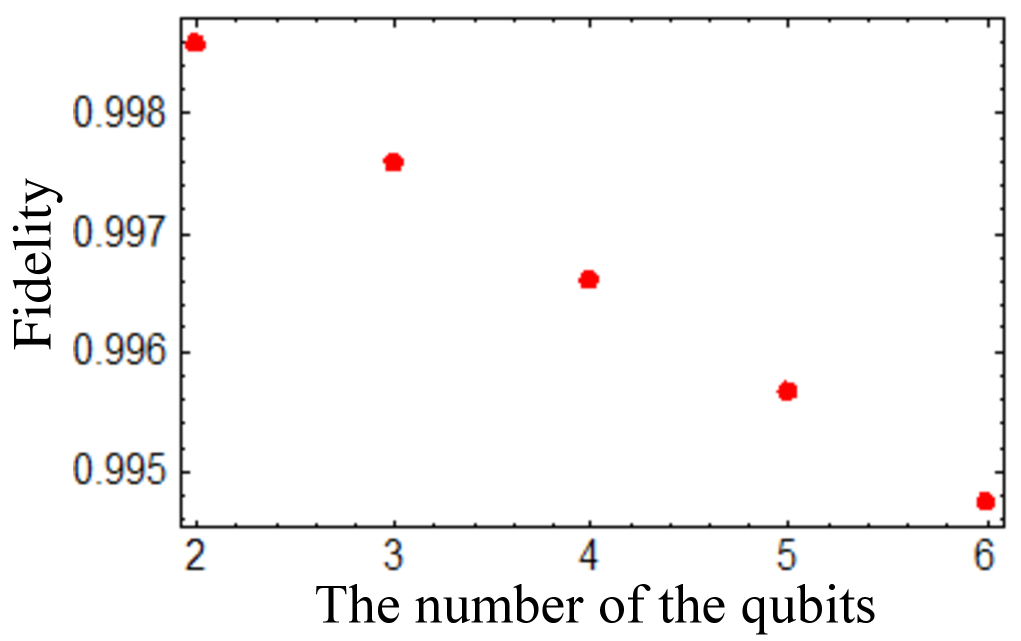}
\caption{The fidelity of the state at the end of the QA plotted against the number of the qubits. Here, we set the parameters as $\lambda /2\pi =1$ GHz, $h/2\pi =0.03$ GHz, $\gamma =0.01$ GHz, $J/2\pi =0.05$ GHz, $T_{{\rm end}}=500$ ns, 
$\omega /2\pi =4.8$ GHz, and $\delta \omega _j /2\pi =1.9(j-1)$ GHz.
}
\label{f2}
\end{figure}

Also, we plot the fidelity of the state at the end of the QA against the number of the qubits as shown in the Fig.\ \ref{f2}. The fidelity decreases almost linearly with the number of the qubits.
Such a linear scaling of the infidelity is typically observed when the independent noise acts on the qubit \cite{hein2005entanglement}. 
So our numerical results indicate that the noise due to the violation of the RWA does not show a spatial correlation but have independent noise properties.


\section{Analytical expression of the violation of the RWA}
In this section, to understand the dynamics shown in Fig.\ \ref{f1}, 
we derive an effective Hamiltonian and will show that the ground state of the effective Hamiltonian is a good approximation
of the exact state numerically obtained by solving the time-dependent Schr\"{o}dinger equation.

We explain the effective dynamics under the periodic driving fields.
In our analysis, we assume that $\gamma ^{-1}$ is much larger than any other time scale  so that the adiabatic condition should be satisfied.
By going to the rotating frame, the Hamiltonian in the Eq. (\ref{achamiltonian}) is written as follows.

\begin{align}
H(t)&= H_{{\rm QA}} + e^{-\gamma ^2t^2}H_{{\rm D'}}(t) +(1-e^{-\gamma ^2t^2})H_{{\rm xy'}}(t),\\
H_{\rm{D'}}&= -\sum_{i=1}^{L}\frac{\Lambda}{2} \left(\cos{[2(\omega+\delta\omega_i)t]} \sigma _x ^{(i)}-\sin{[2(\omega+\delta\omega_i)t]} \sigma _y ^{(i)} \right),\\
H_{\rm{xy'}}&=-\sum_{i,i'=1}^{L}\frac{J_{i,i'}}{2} \left[\cos{[(\delta\omega_i-\delta\omega_{i'})t]} \left(\sigma _x ^{(i)}\sigma _x ^{(i')}+ \sigma _y ^{(i)}\sigma _y ^{(i')}\right)\right.\notag\\
&\qquad \left. +\sin{[(\delta\omega_i-\delta\omega_{i'})t]} \left(\sigma _x ^{(i)}\sigma _y ^{(i')}- \sigma _y ^{(i)}\sigma _x ^{(i')}\right)\right].
\end{align}
We can perform the Magnus expansion \cite{maricq1982application,mananga2011introduction} up to the second order, and obtain the following
\begin{align}
U(t)&={\mathcal{T}}\{e^{-i\int_0^t H(t')dt'}\}\simeq e^{-i(\Omega_1(t)+\Omega_2(t))},\\
\Omega_1(t)&=\int_0^t H(t')dt',\label{firstorder}\\
\Omega_2(t)&=-\frac{i}{2}\int_0^tdt_2\int_0^{t_2}dt_1 [H(t_2),H(t_1)]\label{secondorder}.
\end{align}
where $U(t)$ is the time evolution operator and ${\mathcal{T}}$ denotes the time ordering product. 

\begin{figure}
\includegraphics[width=15.0cm, clip]{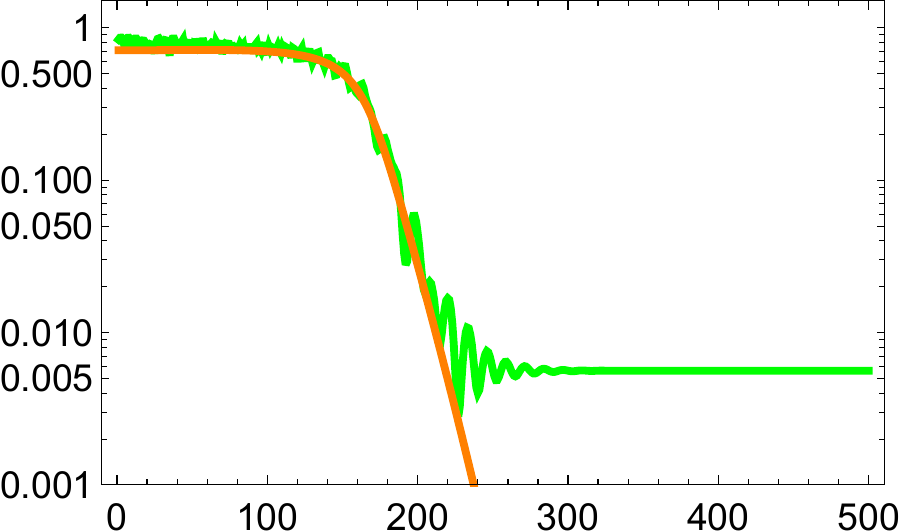}
\caption{The infidelity between $|11\rangle $ and the ground state of the Hamiltonian $H_1+H_2$ plotted against time. Here, we set the same parameters as Fig.\ \ref{f1} (
$\omega /2\pi =2.4$ GHz).
}
\label{f3}
\end{figure}

In order to obtain a simple analytical form, we especially consider the case of $L=2$.
Let us define a period of the oscillation of our system as $T=2\pi n_1/(\omega+\delta\omega_1)=2\pi n_2/(\omega+\delta\omega_2)$ where 
$n_1$ and $n_2$ are positive small integers with $\ T\ll \gamma^{-1}$.
We can define the average Hamiltonian as follows
\begin{align}
H_1&=\frac{1}{T}\Omega_1(T),\label{RWA}\\
H_2&=\frac{1}{T}\Omega_2(T) \label{beyondRWA}.
\end{align}
Starting from a ground state of the Hamiltonian at $t=0$, the system remains in a ground state of the Hamiltonian of $(H_1+H_2)$ for a small $\gamma $.
Note that $H_1$ corresponds to $H_{\rm{QA}}$ (RWA) while $H_2$ corresponds to the violation of the RWA. 
We can calculate $H_2$ as follows.


\begin{align}
H_2&=-\frac{\Lambda^2}{8}e^{-2\gamma ^2t^2}\left(\frac{\sigma_z^{(1)}}{(\omega+\delta\omega_1)}+\frac{\sigma_z^{(2)}}{(\omega+\delta\omega_2)}\right)\notag\\
&+\frac{\Lambda }{4}e^{-\gamma ^2t^2}(1-e^{-\gamma ^2t^2})\left(\frac{h_1\sigma_x^{(1)}}{(\omega+\delta\omega_1)}+\frac{h_2\sigma_x^{(2)}}{(\omega+\delta\omega_2)}+J \left(\frac{\sigma _x ^{(1)}\sigma _z ^{(2)}}{(\omega+\delta\omega_1)}+\frac{\sigma _z ^{(1)}\sigma _x ^{(2)}}{(\omega+\delta\omega_2)}\right)\right)\notag\\
&+\frac{J\Lambda}{2(\delta\omega_1-\delta\omega_{2})}e^{-\gamma ^2t^2}(1-e^{-\gamma ^2t^2})\left(\sigma _x ^{(1)}\sigma _z ^{(2)}-\sigma _z ^{(1)}\sigma _x ^{(2)}\right)\notag\\
&+(1-e^{-\gamma ^2t^2})^2\frac{J^2}{2(\delta\omega_1-\delta\omega_{2})}(\sigma _z ^{(2)}-\sigma _z ^{(1)})
\end{align}
Now, we consider a ground state of the Hamiltonian $H_1+H_2$ at each $t$. 
In Fig.\ \ref{f3}, we plot the infidelity between the state and $|11\rangle $ against time, and compare this result with the infidelity
obtained by numerically solving the time-dependent Schr\"{o}dinger equation. 
There is a good agreement between the results from 
the effective Hamiltonian and that from numerical solutions. 
This shows that our effective Hamiltonian approximately explains the dynamics shown in Fig.\ \ref{f1}, which provides us with the intuition to understand our scheme.

\section{Conclusion}
In conclusion, we propose to implement the QA with CSFQs. 
Although it is difficult to perform the conventional QA by using the CSFQs because of the weak coupling strength and residual flip-flop interactions, we show that a use of the spin-lock based QA can overcome these problems. 
Our numerical simulations show that the spin-lock based QA can be implemented even with the current technology.
\begin{acknowledgments}
This work was supported by Leading Initiative for Excellent Young Researchers MEXT Japan, and is partially supported by MEXT KAKENHI (Grant No. 15H05870) and the New Energy and Industrial Technology Development Organization (NEDO), Japan.

Y.M and H.H equally contributed to this paper.
\end{acknowledgments}




\end{document}